\documentclass[prd,tightenlines,nofootinbib,showpacs,preprintnumbers,superscriptaddress, twocolumn]{revtex4}
\usepackage{amsfonts,amsmath,amssymb,amsthm,bbm,hyperref}
\usepackage{graphicx}
\usepackage{color}
\usepackage[T1]{fontenc}

\newcommand{\be}{\begin{equation}}
\newcommand{\ee}{\end{equation}}
\newcommand{\beq}{\begin{eqnarray}}
\newcommand{\eeq}{\end{eqnarray}}

\begin{document}

\title{Restoration of matter-antimatter symmetry  in the multiverse}
\author{Salvador J. Robles-P\'{e}rez}
\affiliation{Estaci\'{o}n Ecol\'{o}gica de Biocosmolog\'{\i}a, Pedro de Alvarado, 14, 06411-Medell\'{\i}n, Spain.}
\affiliation{Instituto de F\'{\i}sica Fundamental, Consejo Superior de Investigaciones Cient\'{\i}ficas, Serrano 121, 28006 Madrid, Spain,}
\date{\today}

\begin{abstract}
In the multiverse, the universes can be created in entangled pairs with spacetimes that are both expanding in terms of the time variables experienced by internal observers in their particle physics experiments. The time variables of the two universes are related by an antipodal-like symmetry that might explain why there is no antimatter in our universe: at the origin, antimatter is created, by definition and for any observer, in the observer's partner universe. The Euclidean region of the spacetime that separates the two universes acts as a quantum barrier that prevents matter-antimatter from collapse.
\end{abstract}

\maketitle


The multiverse might change some of the preconceptions of the physics of the twentieth  century and explain some of its major puzzles. For instance, in the multiverse the universes can be created in entangled pairs in a composite entangled state that could explain why there is no antimatter in our universe. It would be always created in the observer's partner universe. In order to show it, let us consider a closed DeSitter spacetime with metric
\be
ds^2 = a^2(\eta) ( d\eta^2 - d\Omega_3^2) ,
\ee
where $a(\eta)$ is the scale factor, with $\eta$ being the conformal time, and $d\Omega_3^2$ is the line element on the three dimensional sphere of unit radius. Let us also consider a rescaled conformally coupled massless and charged scalar field, $\chi(\eta)$ (see Refs. \cite{Hartle1983, Gott1998, RP2017c} for the details). The formalism can be easily generalized to other fields\footnote{For a minimally coupled scalar field with mass $m$ one can use the developments made in Refs. \cite{Halliwell1989, Kiefer1992}.}. However, the exactness of the solutions of the conformally coupled scalar field makes the reasoning simpler and clearer.

In that case, the general solution of the wave function of the universe is given, in Everett's terminology\cite{Everett1957}, by a superposition of relative states \cite{Hartle1983, Gott1998, RP2017c}
\be\label{WF01}
\Psi(a,\chi) = \sum_n  \psi_n(a) \phi_n(\chi) b_n  + \psi_n^*(a) \phi^*_n(\chi) c_{-n}^* , 
\ee
where, $\Psi(a,\chi)$, is the function of the universe that encodes all the physical information about the spacetime and the matter fields that propagate therein \cite{Hartle1983},  and $b_n$ and $c_{-n}^*$  in (\ref{WF01}) are two constants that have been written in that particular way for later convenience. The wave function (\ref{WF01}) is the solution of the Wheeler-DeWitt equation \cite{DeWitt1967} that in the case being considered separates \cite{Hartle1983, Gott1998}. The modes $\phi_n(\chi)$ satisfy  the equation of a quantum harmonic oscillator with unit mass and frequency and the modes $\psi_n(a)$ satisfy the following generalized equation of a harmonic oscillator \cite{RP2017c}
\begin{equation}\label{WDW02}
\ddot{\psi}_n(a) + \frac{\dot{\mathcal{M}}}{\mathcal{M}} \dot{\psi}_n(a) + \omega^2_n \psi_n(a) = 0 ,
\end{equation}
where the dot means derivative with respect to the scale factor, i.e. $\dot{\psi} \equiv \frac{\partial \psi}{\partial a}$, $\mathcal{M}\equiv\mathcal{M}(a) = a$ (a particular choice of factor ordering has been taken), and 
\begin{equation}\label{frequency}
\omega_n\equiv\omega_n(a) = \sqrt{ H^2 a^4 - a^2 + 2 E_n} ,
\end{equation}
where, $E_n = n+\frac{1}{2}$, is the energy of the matter field and, $\Lambda \equiv H^2$, is  the cosmological constant of the DeSitter spacetime.

The creation and annihilation of universes in the context of the multiverse is better explained in the so-called third quantization formalism \cite{Caderni1984, McGuian1988, Rubakov1988, Strominger1990, RP2010, RP2013}. It parallels the formalism of a quantum field theory in a curved spacetime but the field to be quantized here is the wave function of the universe and the space where it propagates is the minisuperspace spanned by the variables $\{a, \chi\}$. Thus, the wave function of the universe (\ref{WF01}) is promoted to an operator, $\Psi \rightarrow \hat{\Psi}$, by promoting the constants $b_n$ and $c_{-n}^*$ to creation and annihilation operators of universes, $b_n \rightarrow \hat{b}_n$, and, $c_{-n}^* \rightarrow \hat{c}_{-n}^\dag$. In that case, we can  consider the multiverse as the many particle description of the wave function (\ref{WF01}). It turns out that the universes are created in entangled pairs with opposite momenta \cite{RP2014, RP2017c} in a parallel way as particles are created in pairs with opposite momenta in a homogeneous and isotropic spacetime. In cosmology, however, the momenta conjugated to the scale factor is, $p_a \equiv -  \partial_\eta a$, so it is related to the expansion rate of the spacetime of the universe that is created. Thus, the opposite momenta of the two newborn universes refer to the opposite expansion rates of their spacetimes in terms of a common time variable, i.e. they correspond to the expanding and contracting branches of the wave function of the universe \cite{Hartle1983, Vilenkin1989, Kiefer1992, Kiefer1994}.

As the universe expands the fluctuations of the spacetime diminish, and the momentum distribution of the quantum state of the spacetime is highly peaked around the classical value, $p_a = \omega_n$. The spacetime behaves then classically and  the wave function $\psi_n(a)$ can be approximated by the WKB solutions of (\ref{WDW02}), given by
\be
\psi_n(a) \approx \frac{1}{\sqrt{\mathcal{M} \omega_n}} e^{-i \int^a \omega_n(a') da'} ,
\ee
which are normalized according to, $\psi_n \dot{\psi_n^*} - \dot{\psi_n} \psi_n^* = \frac{2 i }{\mathcal{M}}$. On the other hand, in the same regime the frequency $\omega_n$ in (\ref{frequency}) can be approximated by
\be
\omega_n \approx \omega_{DS} + \frac{1}{\omega_{DS}} ( n + \frac{1}{2} ) ,
\ee
where $\omega_{DS}$ is the square root of the potential of a DeSitter spacetime,
\be
\omega_{DS} = \sqrt{H^2 a^4 - a^2} .
\ee
Then, it turns out that the solution (\ref{WF01}) becomes
\be\label{SC01}
\Psi \approx  \psi_{DS}^+(a)  \phi(\chi, \eta_I)  + \psi_{DS}^-(a)  \phi(\chi^*, \eta_{II}) ,
\ee
where,
\be
\psi_{DS}^+(a) =  \left( \psi_{DS}^-(a)  \right)^* = \frac{1}{\sqrt{\mathcal{M} \omega_{DS}}} e^{-i S_{DS}(a)} ,
\ee
with \cite{Hartle1983}
\be
S_{DS}(a) = \int^a \omega_{DS}(a') da' = \frac{1}{3 H^2} (H^2 a^2 - 1)^\frac{3}{2} ,
\ee
and \cite{RP2017c}
\be\label{SPS01}
\phi(\chi, \eta) = \phi(\chi, a(\eta)) = \sum_n c_n e^{-i (n+\frac{1}{2}) \int\frac{d a}{\omega_{DS}}}  \phi_n(\chi) ,
\ee
is the general solution of the corresponding Schr\"{o}dinger equation of the matter field in the spacetime where it is propagating.

Let us focus on the semiclassical state (\ref{SC01}), which contains the key features about the creation of the universes in entangled pairs. It represents two classical spacetime backgrounds with quantum fields propagating therein. First of all, it is worth noticing that the superposition state $\phi$, given by (\ref{SPS01}), is only a solution of the Wheeler-DeWitt equation in the semiclassical regime. It means that the superposition principle of quantum mechanics turns out to be an emergent feature of the semiclassical description of the universe and is not generally valid. For instance, it is not valid in the spacetime foam \cite{RP2017c}. Of course, the superposition principle of spacetime and matter fields, which is formulated in (\ref{WF01}), is satisfied because it is rooted on the linearity of the Wheeler-DeWitt equation, but the superposition principle of matter fields alone is not generally satisfied and it is only and approximately valid in the semiclassical regime of the spacetime.

In (\ref{SC01}), $\psi_{DS}^\pm(a)$ are the wave functions that represent the quantum states of two branches of the DeSitter spacetime whose momentum distributions are peaked around the two opposite values of the classical momenta, given by $p_a \equiv - a\partial_t a = \pm \omega_{DS}$, where the $+$ sign corresponds to $\psi_{DS}^-(a)$ and the $-$ sign to $\psi_{DS}^+(a)$ (let us notice that, $\hat{p}\psi_{DS}^\pm \equiv - i \partial_a \psi_{DS}^\pm \approx \mp\omega_{DS}\psi_{DS}^\pm$, so $\langle\pm|\hat{p}|\pm\rangle \approx \mp \omega_{DS}$). Then,  $\psi_{DS}^+$ represents the expanding branch of the DeSitter spacetime and  $\psi_{DS}^-$ the contracting branch in terms of a common cosmic time variable, $t = \int a d\eta$. 

However, $t$ is not the time variable measured by internal observers in their particle physics experiments. We have already noticed that the wave function of the universe contains all the physical information about the spacetime and the matter fields. In particular, time and the Schr\"{o}dinger equation are emergent features of the semiclassical description of the universe \cite{Halliwell1989, Hartle1990, Hartle1993, Kiefer1994}. Let us notice that inserting the WKB solutions into the Wheeler-DeWitt equation it is satisfied at first order the Friedmann equation of the DeSitter spacetime background,
\be\label{FE01}
\left( \frac{1}{a} \frac{d a}{d t} \right)^2 = \frac{\omega_{DS}^2}{a^4}  
\ee
At second order, it is also satisfied for a large parent universe like ours \cite{Hartle1990}
\be\label{SCH01}
\mp 2 i \hbar \omega_{DS} \frac{\partial \phi(\chi)}{\partial a} = \left( - \hbar^2 \frac{\partial^2}{\partial \chi^2} + \chi^2 \right) \phi(\chi) ,
\ee
which is the time dependent Schr\"{o}dinger equation for the field $\chi(t)$ provided that the time variable $t$ has been defined according to (\ref{FE01}), i.e.
\be
\frac{\partial }{\partial t} = \pm \ \frac{\omega_{DS}}{a} \frac{\partial}{\partial a } ,
\ee
where the $+$ sign corresponds to the expanding branch of the universe and the $-$ sign to the contracting branch. Therefore, from the quantum cosmological standpoint time arises as an emergent feature of the semiclassical regime of the wave function of the universe. An observer inhabiting each branch of the entangled pair defines his or her time variable from the experiments of particle physics that are governed by the Schr\"{o}dinger equation (\ref{SCH01}) at a particular moment of the cosmic expansion, $a_0$. Then, the inhabitants of the expanding branch define their time variable as, $t_I \equiv t$, and the inhabitants of the contracting branch\footnote{Contracting with respect to the time variable $t$.} define their time variable as, $t_{II} \equiv - t$, which is the time variable they experience in their particle physics experiments, i.e. it is the time variable measured by their actual clocks. Thus, from the point of view of the time variable experienced by an internal observer, both wave functions $\psi_{DS}^+(a)$ and $\psi_{DS}^-(a)$ describe an expanding universe.

Therefore, the  state (\ref{SC01}) represents the entangled state between two spacetimes, which are both expanding universes in terms of the time variable experienced by internal observers. In the each semiclassical branch of the DeSitter spacetime the superposition principle of quantum mechanics is satisfied (at least to order $\hbar^1$) and the wave function $ \phi$ is the superposition state (\ref{SPS01}) of two fields, $\chi$ and $\chi^*$, which are propagating respectively in the two patches of the DeSitter spacetime, with conformal time variables $\eta_I$ and $\eta_{II}$ given respectively by
\beq
\eta_I &=& \int^a \frac{d t_I}{a} = \int^a \frac{d t}{a}  = \int^a \frac{d a' }{\omega_{DS}(a')} , \\
\eta_{II} &=& \int^a \frac{d t_{II}}{a} = - \int^a \frac{d t}{a} =  - \int^a \frac{d a' }{\omega_{DS}(a')} . 
\eeq
It means that the time variables of the two universes are related by an antipodal like symmetry \cite{Linde1988, Linde1991}, $t_I = - t_{II}$. Let us also notice that the complex conjugated of (\ref{SCH01}) becomes 
\be\label{SCH02}
\pm 2 i \hbar \omega_{DS}  \frac{\partial \phi(\chi^*)}{\partial a}= \left( - \hbar^2 \frac{\partial^2}{\partial {\chi^*}^2} + {\chi^*}^2 \right) \phi(\chi^*) ,
\ee
where, $\phi(\chi^*) = \phi^*(\chi)$, so according to the entangled state (\ref{SC01}) matter is created, by definition, in the observer's universe and antimatter is always created in the partner universe, and this is so for the observers of both single universes. Hence, the matter of our universe would be the antimatter of our hypothetical partner universe and the missing antimatter of our universe would be the matter of our partner universe. The matter-antimatter asymmetry would then be restored in the context of a multiverse in which the universes are created in entangled pairs stem from the existence of double Euclidean instantons \cite{RP2014, RP2017c}. The Euclidean region of the spacetime that separates the Lorentzian patches of the spacetime acts as a quantum barrier and prevents matter-antimatter from collapse.

The customary development of quantum field theory in the DeSitter background would follow as usual \cite{Mamaev1976, Birrell1982, Mukhanov2007} except that the mode expansion of the charged scalar field becomes
\be
\chi = \int d\mu(k)   \left( \zeta_k(\textbf{x}_I) v_k^*(\eta_I)  \hat{a}_k + \zeta^*_k(\textbf{x}_{II}) v_k(\eta_{II}) \hat{b}^\dag_{-k}    \right) ,
\ee
where $d\mu(k)$ is the measure of the Fourier space \cite{Birrell1982}, $\zeta_k(\textbf{x})$ are the eigenfunctions of the three dimensional Laplacian in the spacetime of each single universe,  $\hat{a}_k^\dag$ and $\hat{a}_k$ are in our universe\footnote{In the partner universe it is the other way around, $\hat{a}_k^\dag$ and $\hat{a}_k$ are the creation and annihilation of antiparticles, and $\hat{b}_{-k}^\dag$ and $\hat{b}_{-k}$ are those for particles.} the creation and annihilation of particles, and $\hat{b}_{-k}^\dag$ and $\hat{b}_{-k}$ are those for antiparticles  \cite{Mamaev1976}.

If the boundary condition for the field is global and such that the field is in the invariant vacuum state \cite{RP2017d}, then, the quantum states of the particles of the universe would be given by the reduced density matrix that is obtained by tracing out from the composite vacuum state the degrees of freedom of the particles of the partner universe. It turns out that the particles of each single universe are distributed following a quasi-thermal state given by \cite{RP2012, RP2017d}
\be\label{RHO01}
\rho =  \prod_\textbf{k} \frac{1}{Z_k} \sum_n e^{-\frac{1}{T_k} (n+\frac{1}{2})} | n_{\textbf{k}} \rangle \langle n_{\textbf{k}}| ,
\ee
where, $Z^{-1} = 2 \sinh\frac{1}{2 T_k}$, is the partition function, with a time dependent temperature given by
\be
T_k^{-1} \equiv T_k^{-1}(\eta) = \ln\left( 1+\frac{1}{N_k}\right) ,
\ee
with, $N_k = |\nu_k|^2$, being the number of particles in the universe, and $\nu_k$ is the Bogolyubov coefficient that relates the invariant representation of the composite vacuum of the fields in the two universes with the representation that describes the observable particles in one of the single universes (see, Ref. \cite{RP2017d} for the details). In the case that this is given by the instantaneous diagonal representation of the Hamiltonian of the field, it yields for the semiclassical branch of a DeSitter universe \cite{RP2017d}
\be
\nu_k \approx \frac{\dot{a}^2}{4 k^2} = \frac{H^2}{4 k_\text{ph}^2} ,
\ee
where, $\dot{a} \equiv \frac{\partial a}{\partial t}$, and, $k_\text{ph} \equiv \frac{k}{a(t)}$, is the physical wavelength of the field. The field of each single universe is not in the vacuum state but in the quasi-thermal state given by (\ref{RHO01}). It affects the expected value of the amplitude of fluctuations of the matter fields during the inflationary stage of the universes because these do not have to be computed for the vacuum state but for the thermal state (\ref{RHO01}). The way in which this modifies the theoretical estimations of the power spectrum of the CMB for a more accurate model of our universe is something that deserves a deeper analysis but it might provide us with observational imprints of the existence of a partner universe, giving experimental support to the whole multiverse proposal and solving some of the major existing puzzles of contemporary physics.


\bibliographystyle{apsrev}

\end{document}